\documentclass{JINST}

\usepackage{bm}

\title{Method of reconstructing a moving pulse}

\author{Stephen~J.~Howard\thanks{Corresponding
author.}, Robert~D.~Horton, David~Q.~Hwang, Russell~W.~Evans, Samuel~J.~Brockington,  and Jeffrey~Johnson\\
UC Davis Department of Applied Science,\\
Livermore, CA, 94551, USA\\
E-mail: \email{showard@ucdavis.edu}}

\abstract{We present a  method of analyzing a set of $N$ time signals $f_i(t)$ that consist of local measurements of the same physical observable taken at $N$ sequential locations $Z_i$ along the length of an experimental device. The result is an algorithm for reconstructing an approximation $F(z,t)$ of the field $f(z,t)$ in the inaccessible regions between the points of measurement. We also explore the conditions needed for this approximation to hold, and test the algorithm under a variety of conditions. We apply this method to analyze the magnetic field measurements taken on the Compact Toroid Injection eXperiment (CTIX) plasma accelerator; providing a direct means of visualizing experimental data, quantifying global properties, and benchmarking simulation.}


\begin{document}

\section{Introduction}

Here we present a new method to reconstruct an accurate estimation of the instantaneous spatial dependence of an experimentally measurable field, using the signals from a small set of spatially distributed time-domain measurements. Such a reconstruction is only possible if the field is being transported past the sequence of measurement locations at a speed that is comparable to the rate at which field growth or decay is occurring within individual fluid elements. In this paper the term ``field'' will refer to either a  variable property of a fluid medium (such as density or pressure), or to an independent quantity (such as an electromagnetic field), which may or may not couple to a physical fluid. 

We call this method \emph{co-moving interpolation} because it follows the trajectories of fluid elements through the system and performs an interpolation of the field quantity using the measured values at the points where a trajectory crosses the probe locations, doing so along the path of each trajectory. The algorithm for this method is described in detail in sections \ref{LIformula} and \ref{TrajAlgor}.

The development of this method was motivated by data from the Compact Toroid Injection eXperiment (CTIX), \cite{mclean:ctix}, \cite{DavisThesis}  a plasma accelerator device that generates high speed magnetized plasma rings called compact toroids ($\mathrm{v}\ \sim$ 200 km/s, B $\sim$ 10 kGauss), primarily for the purpose of refueling a tokamak fusion reactor. This interpolation method has been particularly helpful in resolving the issue of the magnetic geometry of the compact toroid plasma, which determines how the CT will interact with the magnetic field of the reactor as it deposits fuel at its interior. In section \ref{SeqMeas} we will present an overview of how this interpolation method is used on CTIX, while in section \ref{VelCTIX} we will look at the particular method of velocity estimation that works well for our system.

This interpolation relies on first making an estimation of the velocity field throughout the system, usually based on the apparent time-of-flight kinematics of any traveling pulses. We treat the velocity estimation as an independent problem that is not necessary to fully explore in order to understand the key features of the co-moving interpolation algorithm.
However, in section \ref{BShock} we will examine an analytic model of shock propagation as a way to compare an exact solution against three cases of interpolation using different velocity fields. This comparison illustrates an important distinction between the fluid velocity field and the optimal reconstruction velocity field.

Lastly, in section \ref{ErrAnalysis} we will examine two different approaches to find upper bounds on the intrinsic errors of this method and show how they scale with the system parameters, (e.g., number of probes, flow velocity, probe separation). We demonstrate the application of this method to facilitate the direct comparison between simulation, analytic models, and experiment.

\section{Sequential measurements of a moving pulse}\label{SeqMeas}
The motivation for developing this type of interpolation originates with the task of trying to interpret experimental time-domain signals. 
Effective implementations of this algorithm can produce a quantitative analysis of the accelerator dynamics that significantly improves upon previous methods. (Compare to \cite{mclean:ctix}, \cite{woodall:jap}, \cite{Bhuyan:1}).

\begin{figure}[!ht] 
\begin{center}
\includegraphics[width=\textwidth]{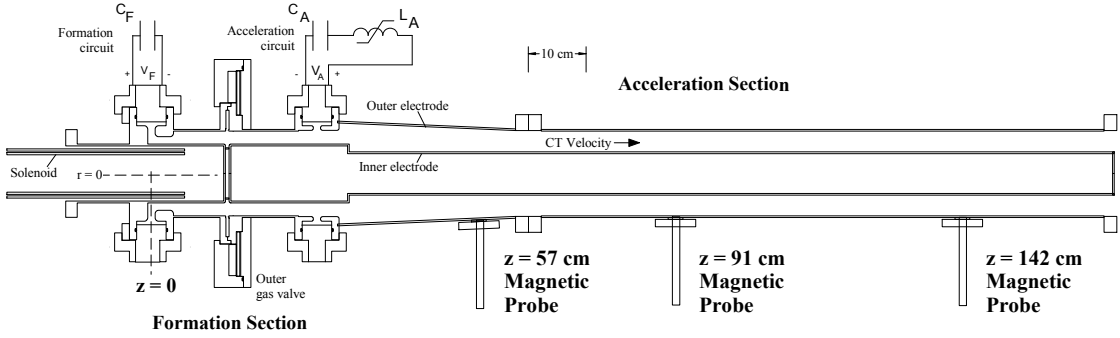}
\caption{Schematic of the CTIX plasma accelerator showing the coaxial inner and outer electrodes, the formation and accelerator circuits, and the location of the magnetic probes. The co-moving interpolation method has been helpful in analyzing the magnetic data for this system.} \label{fig:CTIXschematic} 
\end{center}
\end{figure}

On CTIX our interest is in reconstructing the approximate spatial dependence of the magnetic field within the plasma, using data collected from three magnetic field probes that measure the edge magnetic field of an accelerated plasma pulse. 
Sample data is presented below in figure \ref{fig:BzTrio59842} showing the time dependence of the magnetic field measured by probes. 
In Figure \ref{fig:CTIXschematic} we see a schematic of the experimental arrangement in which these measurements were taken.

\begin{figure}[!ht] 
\begin{center}
\includegraphics[width=.75\textwidth]{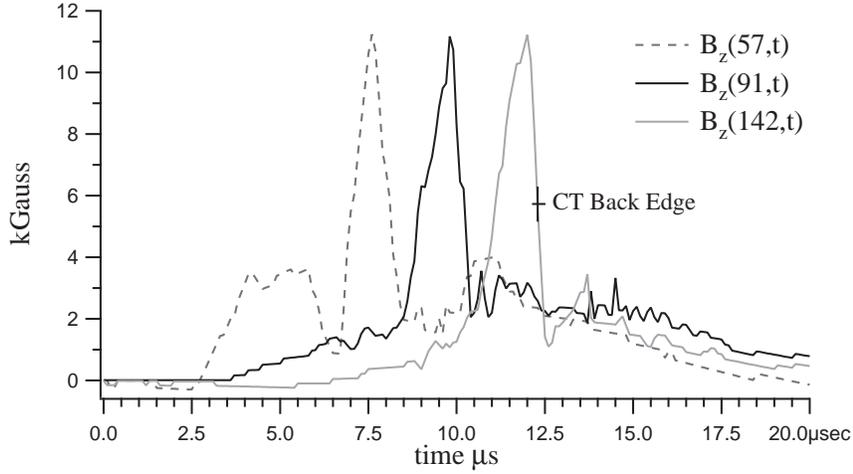} 
\caption{Example of CTIX magnetic field data, three measurements of the axial component of the edge magnetic field $B_z(t)$ taken within the CTIX plasma accelerator at positions z = 57, 91, 142 cm.}
\label{fig:BzTrio59842}
\end{center}
\end{figure}

It is interesting to note that this general concept is also used in video decompression and key frame animation when trying to interpolate in time between original frames of a video image \cite{Chahine:dvc},  \cite{Burtnyk:1}, although the mathematical formalism and resulting algorithm are not applicable to analyzing the data from a set of diagnostic probes, and a new formulation is required. 

Any implementation of this method must begin by making an accurate estimate of the fluid velocity field for all $(z,t)$ in the system. 
Velocity estimation is a difficult problem in general, \cite{Stefani:ip} but many experimental systems have some simplifying symmetry or property that allows the flow velocity to be inferred from probe measurements. In the case of CTIX, the simplifying property used is the approximation of constant acceleration, as discussed in section \ref{VelCTIX}.

Fluid elements being tracked may not necessarily correspond to any physical fluid with mass density. These may correspond to the trajectories inferred from either the group velocity or phase velocity of the wave pattern that is being reconstructed.
The interpolation algorithm takes this estimated velocity field as an input and uses it to create fluid element trajectories, on which it computes a superposition of the adjacent real data to determine the interpolated value of the observable along each trajectory. 

\section{Co-moving interpolation formula}\label{LIformula}
This key formula defines the interpolated value $F(z,t)$ of the observable $f(z,t)$ at position $z$ between the $ith$ and \mbox{$(i+1)th$} real probes, using a superposition of neighboring probe signals evaluated at retarded and advanced times:
\begin{equation}
F(z,t) = w_i(z,t) {f_i}(\tau_i(z,t))\ +\  (1-w_i(z,t)) {f_{i+1}}(\tau_{i+1}(z,t)), \label{intfzt}
\end{equation}
where $w_i$ is a non-negative weighting factor $(\leq 1)$ that has boundary conditions \mbox{$w_i(Z_i,t)\ \equiv\ 1$}, \mbox{$w_i(Z_{i+1},t) \equiv 0$}. 
The function $f_i(t)$ is the time signal of the $ith$ real probe $f_i(t) = f(Z_i,t)$ where $Z_i$ is the position of the $ith$ probe, $i = 1,2,... N$. The time shift function $\tau_i(z,t)$ records the time that the fluid element at $(z,t)$ crosses through the $z = Z_i$ position. For a fluid that is moving in the positive $z$ direction, a given fluid element first passes by the lower probe at $Z_i$, then passes through the value of $z$ under consideration for interpolation, and finally passes the upper probe at $Z_{i+1}$. For all $z$ in the interval $[Z_i, Z_{i+1}]$, we have $\tau_i(z,t) \leq t \leq \tau_{i+1}(z,t)$.
The time shifts must equal the identity at the probe locations, $\tau_i(Z_i,t) \equiv t$ for all $t$.

Although in general the weighting function $w_i$ can depend on $z$ and $t$, an efficient time-independent method is to interpolate linearly in space. 
\begin{equation}
w_i(z,t) = (z - Z_i) / (Z_{i+1} - Z_i) \label{equ:zweight}.
\end{equation} 
Alternatively, a linear interpolation in the time dimension could be performed using 
\begin{equation}
w_i(z,t) = (t - \tau_i(z,t)) / (\tau_{i+1}(z,t) - \tau_i(z,t)). \label{equ:tweight}
\end{equation} 
This is slightly more computationally expensive than the time-independent method due to the look-up time of $\tau_i(z,t)$ compared to the constant $Z_i$,  but it does a better job at matching rates of change of observables that do not stay constant along the fluid trajectories. 
Higher-order polynomial interpolation along the trajectories is also possible, but may be unnecessarily complicated given the good results of the methods that have been implemented so far using simple linear weighting functions.

\section{Trajectory  algorithm for $\bm{\tau_i(z,t)}$}\label{TrajAlgor}
For continuous, integrable velocity fields the trajectory of an individual fluid element given by $z = z_{\mu}(t)$ is related to the velocity $\mathrm{v}(z,t)$ according to
\begin{equation}
\frac{d z_{\mu}(t)}{d t} = \mathrm{v}(z_{\mu}(t),t) = \mathrm{v}_{\mu}(t). \label{Diff_zmu_vel}
\end{equation} 

Here $\mu$ is a Lagrangian coordinate that uniquely labels the fluid elements, which can be defined as the position of the fluid element at some initial time $t_0$, by $z_{\mu}(t_0) = \mu$. 
A second order Runge-Kutta algorithm \cite{BurdenFairs:Numerical} is used to compute each trajectory from the given velocity field $\mathrm{v}(z,t)$ and find the time that it crosses the neighboring probe locations. For each probe position $Z_i$ there is a distinct sub-domain of grid points $\Omega_i = [Z_{i-1},Z_{i+1}]\times[0,N_t]$ over which the crossing time $\tau_i(z,t)$ needs to be evaluated. 
The trajectory waveform is composed of a time-coordinate array $\{t_j\}$ and a space-coordinate array $\{z_j\}$. The magnitude of the timestep $|h|$ is a small arbitrary constant that is fixed before runtime. 

\begin{figure}[!ht]
\begin{center}
\includegraphics[width=.75\textwidth]{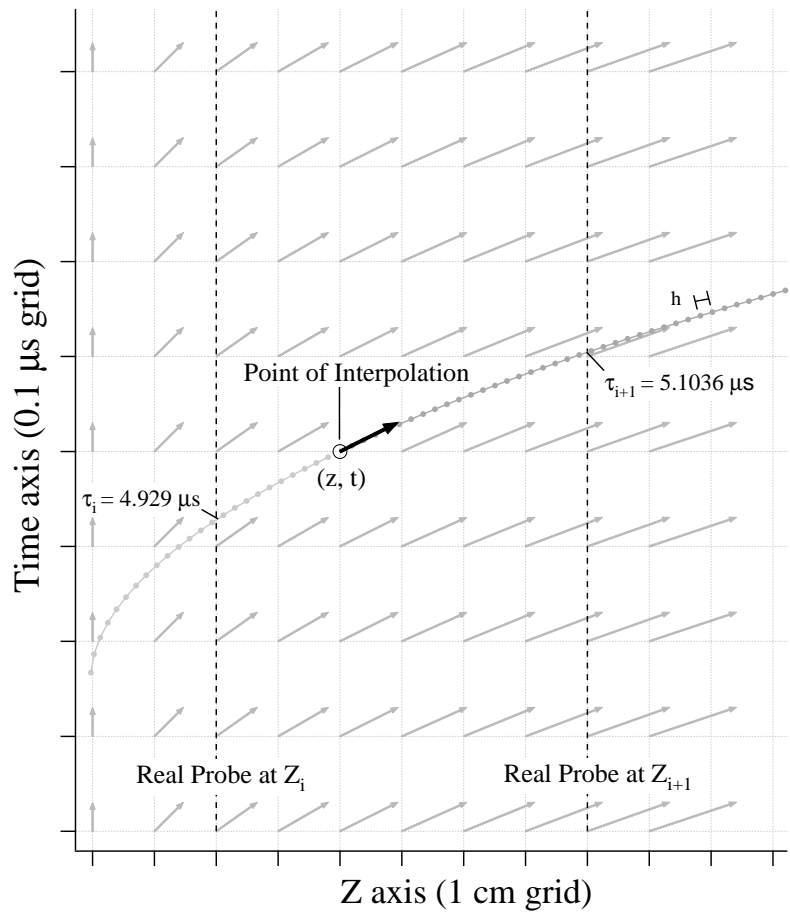}
\caption{The estimated velocity field on the (z,t) domain. The trajectory of a fluid element is shown as the grey curve as it passes the probe at position at $Z_i$ and then the probe at $Z_{i+1}$.} 
\label{fig:VelTraj}
\end{center}
\end{figure}

Integration of (\ref{Diff_zmu_vel}) is accomplished with the following algorithm.
\paragraph*{Outer Loop} For $i = 1, 2, ... N$, set $Z_i$ equal to the $ith$ real probe location and evaluate the following \emph{Steps 1-4} for every $(z,t)$ pair in the $ith$ sub-domain $\Omega_i$.
\paragraph*{Inner Loop: Step 1} Start the trajectory at $t_0 = t, \ z_0 = z$. Set the sign of the timestep $h$ depending on the direction toward the probe at $Z_i$: $h > 0\ \mbox{if}\ z < Z_i$, $h < 0\ \mbox{if}\ z > Z_i$.
\paragraph*{Step 2} Find the next trajectory point using the Runge-Kutta midpoint method
\begin{eqnarray}
	t_{j+1}  &=& t_{j} + h \\
	z_{j+1}  &=& z_{j} + h\ \mathrm{v}\!\left(z_j + \frac{h}{2}\mathrm{v}(z_j,t_j),\ t_j + \frac{h}{2} \right)
\end{eqnarray}
\paragraph*{Step 3} Repeat \emph{Step 2} (iterating $j=j+1$) until the interval between $z_j$ and $z_{j+1}$ contains the probe location $Z_{i}$. After crossing the probe location, the final point added to the trajectory waveform is $(z_{j+1}, t_{j+1})$.
\paragraph*{Step 4} Find the crossing time by interpolating across the interval $[z_j, z_{j+1}]$, such that $\tau_i(z,t)$ is in the interval $[t_j,t_{j+1}]$. For a linear interpolation of the crossing time, set 
\begin{equation}
\tau_i(z,t) = t_j + h \cdot \frac{Z_i - z_j}{z_{j+1} - z_j}. 
\end{equation}
Refer to figure \ref{fig:VelTraj} for a visual depiction of \emph{Steps 1-4}. When our implementation of this method is run on a 1.8 GHz Pentium 4 computer it yields a runtime of about \mbox{4 $\mu s$} per timestep on an individual trajectory calculation. The local truncation error of the trajectory algorithm is $\mathit{O}(h^2)$.

Once the crossing times have been tabulated in $\tau_i(z,t)$ for $i = 1,2,...N$, the weighting functions $w_i(z,t)$ can be evaluated, and the co-moving interpolation formula (\ref{intfzt}) can be applied for all $(z,t)$ in the full domain.

\section{Frozen-in approximation}
The concept behind formula (\ref{intfzt}) is the notion that if an ideal fluid has zero substantial derivative for a given observable $f$, 
\[\frac{\partial f}{\partial t} + \mathrm{v} \frac{\partial f }{\partial z} \equiv 0\]
then $f$ is purely convected along by the velocity field, or ``frozen'' into the fluid, and the velocity field alone contains all of the dynamical information of the system. The frozen-in approximation $F_i(z,t)$ of $f(z,t)$ about the measurement location $Z_i$ is simply
\begin{equation}
	F_i(z,t) = f_i(\tau_i(z,t)), \label{frozen_approx}
\end{equation}
where $f_i(t) = f(Z_i, t)$ and $\tau_i(z,t)$ is the time that fluid trajectories cross $Z_i$ as determined from a known velocity field.
This is exactly the $N = 1$ case of the general interpolation formula (\ref{intfzt}), with $w_i(z,t) \equiv 1,$ for all $z,t$. 

We find that for ideal fields that are perfectly frozen into the fluid, the reconstructed field can be made to agree with the original model field arbitrarily well. However for real fields that grow or decay as they convect the $N = 1$ interpolation will have an error that can increase without bounds as the distance away from the measurement location $Z_i$ increases. 

The full interpolation (\ref{intfzt}) with multiple probes will improve the reconstruction provided that enough probes are used to guarantee that the characteristic time $\zeta$ of either growth or decay of the field is longer than the transit time between sequential probes. This requires that the growth or decay rate be constrained by:
\begin{equation}
\frac{1}{\zeta}\ \equiv\ 
\frac{1}{\langle f \rangle}\left(\frac{\partial f}{\partial t} + \mathrm{v} \frac{\partial f }{\partial z}\right)\ \leq\ \frac{\langle \mathrm{v} \rangle}{\Delta L_{probe}}, \label{equ:decaytime}
\end{equation}
where $\langle f \rangle$ is the average value of the field, $\langle \mathrm{v} \rangle$ is the average velocity, and $\Delta L_{probe}$ is the  distance between adjacent probes; for $N$ uniformly spaced probes, $\Delta L_{probe} = L/(N+1)$, where $L$ is the total length of the system. 
If (\ref{equ:decaytime}) is not satisfied then more probes are needed to make the co-moving interpolation work well on the given system.

\section{Reconstruction of test case: Burgers' shock} \label{BShock}
 
A simple nonlinear one-dimensional model of shock propagation is given by Burgers' equation \cite{whitham:waves} for the characteristic velocity $c(z,t)$ in a viscous fluid, 
\begin{equation}
	\frac{\partial c}{\partial t} + c \frac{\partial c}{\partial z} = \nu \frac{\partial^2 c}{\partial z^2}, \label{BurgersEqu}
\end{equation}
where $\nu$ is the kinematic viscosity and provides a diffusion mechanism for smoothing out discontinuities near the shock front.
This has a steady-state shock solution of the form
\begin{equation}
	c(z,t) = c_1 + (c_2 - c_1)\left[1 + \exp{\left(\frac{c_2 - c_1}{2 \nu}(z - U t)\right)}\right]^{-1}, \label{BurgerShock}
\end{equation}
where the constants $c_1 <c_2$ give, respectively, the characteristic speed of wave propagation far ahead, and far behind the shock front, while $U$ is the speed of the shock front itself. In the case considered here $U$ is the average of the two speeds, $U = \frac{1}{2}(c_1 + c_2)$. The shock thickness is determined by the diffusive term, with an \emph{e}-folding thickness $L_{shock} = 4 \nu /(c_2 - c_1)$. 
\begin{figure}[!ht] 
\begin{center}
\includegraphics[width=\textwidth]{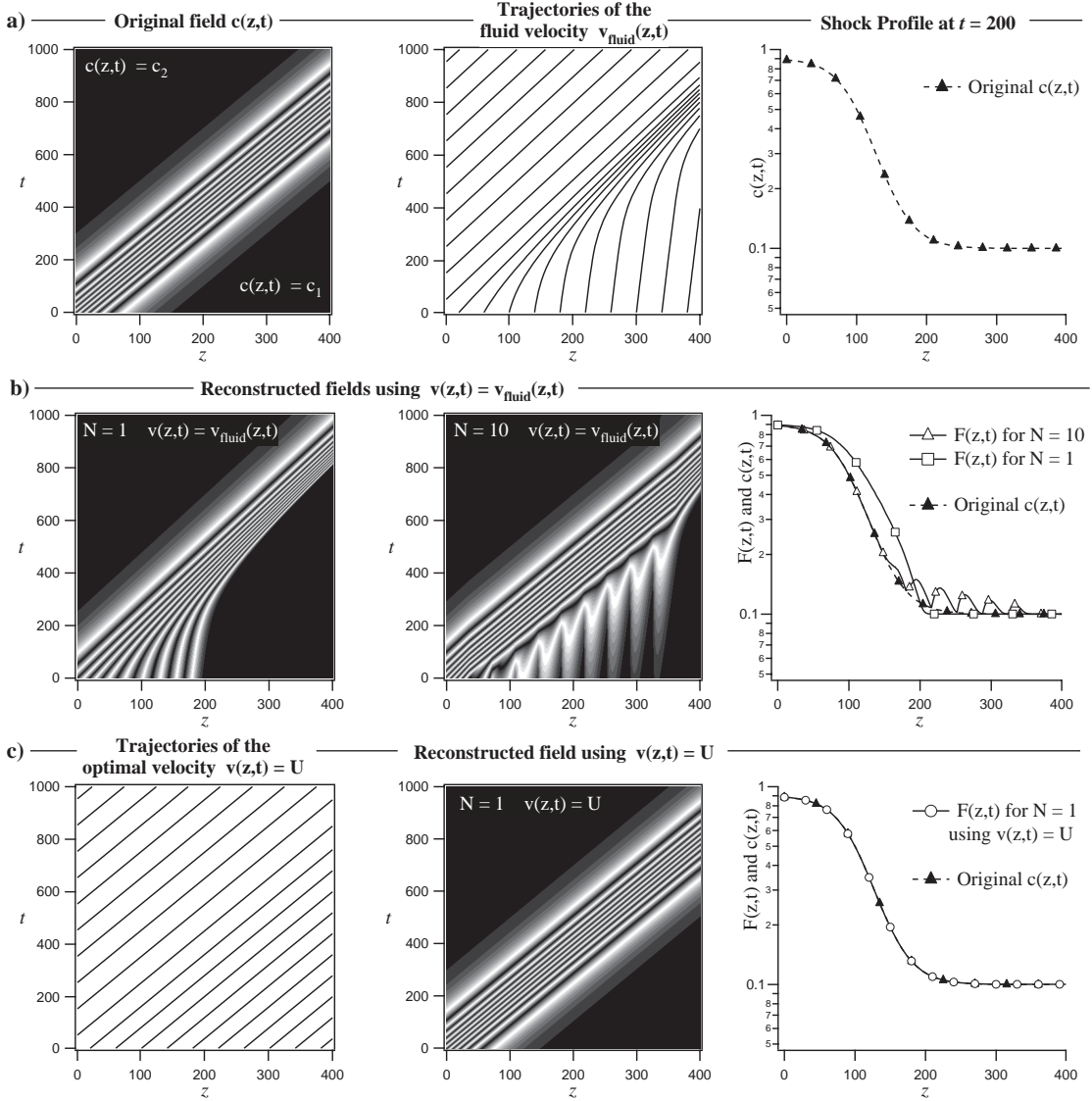}
\caption{\textbf{Burgers' shock wave tests.} a) Original shock wave $c(z,t)$ represented in a cyclic grey-scale to emphasize gradients (left). Trajectories of fluid velocity are shown (middle). The steady-state shock profile, (right) shown at time $t = 200$. b) The co-moving interpolation using $\mathrm{v}_{fluid}(z,t)$ as the velocity field results in a fan-out backwards in time, shown with $N = 1$ (left) and with $N = 10$ (middle). We then compare the reconstructed profiles with the original (right). c) Reconstruction with the optimal velocity $\mathrm{v}(z,t) = U$ using only one probe $(N = 1)$.}
\label{fig:BurgersShock}
\end{center}
\end{figure}

For a density flux given by $\rho \mathrm{v}_{f} = \mathit{q} = Q(\rho) - \nu \partial \rho/\partial z$, the characteristic velocity is defined as $c(\rho) = Q'(\rho)$. In the case of simple model of flux with a parabolic dependence on $\rho$, $Q(\rho) = \alpha \rho^2$, the characteristic velocity depends linearly on density $c(z,t) = 2 \alpha \rho(z,t)$. The fluid velocity for this system is 
\begin{equation}
\mathrm{v}_f\ =\ \frac{\mathit{q}}{\rho}\ =\ 	\alpha \rho - \frac{\nu\ \partial \rho}{\rho\ \partial z}\ =\ \frac{\mathit{c}}{2}  - \frac{\nu\ \partial \mathit{c}}{\mathit{c}\ \partial z}.
\end{equation}

This model is interesting as it applies to this interpolation, because (\ref{BurgersEqu}) equates the substantial derivative of the $c(z,t)$ with $\nu \partial^2 c/\partial z^2$, and so for $\nu \neq 0$ trajectories that follow the motion of wave elements will pass through changing values of $c(z,t)$. This means that the field $c(z,t)$ is not ``frozen'' into the velocity field $c(z,t)$. As it flows, it changes at a rate defined by the viscosity and the second derivative of $c$. The same is true for the fluid velocity $\mathrm{v}_f$ and the density that is transported by it.

 If we use $\mathrm{v}_f(z,t)$ as the velocity field for the co-moving interpolation for small $N \sim 3$, the reconstructed field compares poorly to the \emph{a priori} model. The source of the problem is that the convergence of the actual fluid trajectories toward the shock front results in a ``fan-out'' going backwards in time, shown in figure \ref{fig:BurgersShock}. This problem is mitigated if  more probes are added to lessen the span of the interpolation, but the problem is fundamental.
 
We can quantitatively compare the shock wave field $c(z,t)$ given by (\ref{BurgerShock}) against the results of co-moving interpolation for three different cases on the basis of absolute value of relative error $|R_{err}| = \left|(F(z,t) - c(z,t))/c(z,t)\right|$. 

\vspace{10pt}
\noindent\textbf{Case 1.} $\mathrm{v}(z,t) = \mathrm{v}_f(z,t)$, $N = 1$: The interpolation yields $\max|R_{err}|= 2.41$ and a mean $\langle|R_{err}|\rangle = 0.143$. This large error is understandable given the fact that $c(z,t)$ is not frozen into $\mathrm{v}_f(z,t)$.
See Figure \ref{fig:BurgersShock} b) left.

\vspace{10pt}
\noindent\textbf{Case 2.} $\mathrm{v}(z,t) = \mathrm{v}_f(z,t)$, $N = 10$: With ten probes the result is better with $\max|R_{err}|= 0.492$ and a mean $\langle|R_{err}|\rangle = 0.032$. The error is largest halfway between probes, which results in a oscillatory behavior in the reconstructed shock profile, shown in figure \ref{fig:BurgersShock} b) middle.

\vspace{10pt}
\noindent\textbf{Case 3.} $\mathrm{v}(z,t) = U$, $N = 1$: The interpolation is near-optimal with maximum $\max|R_{err}|= 2.6 \times 10^{-3}$ and a mean of $\langle|R_{err}|\rangle = 2.0 \times 10^{-4}$, which is at the level of the trajectory integration error. See Figure \ref{fig:BurgersShock} c).

The conclusion to take from this test comparison is that the translational velocity $U$ of the steady-state wave pattern results in a better reconstruction than when the actual fluid velocity field is used. This distinction is most important when the fluid is not ideal and viscous diffusion is present.  When we use the optimal reconstruction velocity $\mathrm{v}(z,t) = U$, then the co-moving interpolation is optimal even with a single probe measurement ($N = 1$). 
In general, if we are trying to reconstruct the observable $f(z,t)$ we would get the best results if we could perform the interpolation along trajectories that are the level sets of $f$. In principle this could be done by computing 
 \[\mathrm{v}(z,t) =  - \frac{\partial f /\partial t}{\partial f /\partial z}. \]  
In the real problem we only have the $N$ probe signals to work with, and the apparent velocity that can be inferred from them. As we have seen, this apparent velocity of any steady wave pattern is very close to the optimal velocity field even if the actual fluid velocity is known by some other means.

\section{Velocity estimation for the CTIX system}\label{VelCTIX}
In systems with freely accelerating flows, a velocity estimation method similar to the one used on CTIX is likely to have some success. 
We estimate the flow field on CTIX by tracking one distinct feature as it travels down the accelerator, and then apply the same relative kinematics to the rest of the fluid elements in the system.  The most stable feature on the CTIX waveforms is the back edge of CT; see Figure \ref{fig:BzTrio59842}. This is the junction point between the CT and the pushing field, and we can reliably define the arrival of this point as being the time when the $B_z$\ signal of a given probe crosses the half-maximum level on the trailing edge. The high accuracy of this method is due to the reproducibility of the steep slope that occurs at the CT back edge.
The crossing times at which the CT back edge passes by the three accelerator probes 
serve as input data for a simple kinematic analysis of the average velocities, and overall acceleration.
This yields a constant-acceleration fit for the trajectory of the CT back edge.
\[ z_{CT}(t) =  z_0 + \mathrm{v}_0 t + \frac{1}{2} \mathrm{a} {t^2}. \]
For positive values of acceleration there will always be a  minimum z-position,  $z_{min} \leq z_{CT}(t)$, defined by
\[ z_{min} = z_{0} - \frac{\mathrm{v}_0^2}{2\mathrm{a}}. \]
This is the point on the trajectory where $\mathrm{v}(t) =0$. However, for early times before the fluid element has reached $z_{min}$, according to this description, the CT would have a negative velocity which is unphysical for our system. Instead we will only apply the uniformly accelerated model to the region $z > z_{min}$, and during times of positive velocity. 
If we work with a simple model in which all the fluid elements have the same acceleration throughout time and space, they will all have the same $z_{min}$, but they would pass through it at different times. Based on this, we arrive at a velocity field of the form 
\begin{equation}
	\mathrm{v}(z,t) = \sqrt{2 \mathrm{a}(z - z_{min})}
\end{equation}
Notice that this depends on space but not on time. 
When $z_{min}> 0$ we need a different way to handle the velocity in the region $0 < z < z_{min}$. Typically $z_{min} < 57$cm, which is a region where velocity measurements are unavailable.  One solution is to truncate the domain, and only work where the probe data implies a velocity field.  This is fine when possible, but certain applications of this method require making a velocity estimate over the entire domain, (such as the interpretation of a Doppler-shift measurement along an axial chord). 
\begin{figure}[!hb] 
\begin{center}
\includegraphics[width=\textwidth]{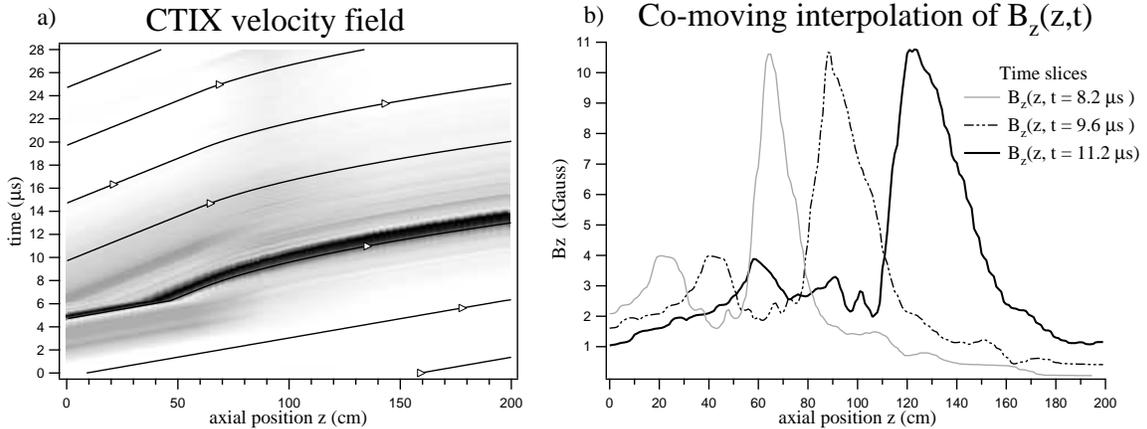}
\caption{Example of co-moving interpolation of magnetic field data. (a) The trajectories of the estimated velocity field using time-of-flight kinematics from probe pulses. The image of the interpolated $B_z(z,t)$ field is overlaid in a grey-scale (white $= 0$, dark $> 0$) to illustrate where the CT is in the system. A constant velocity field is used before the accelerator current fully turns on at about 6 $\mu s$. (b) The reconstructed axial dependence of the magnetic field at three moments in time, $t = 8.2,\ 9.6,\ 11.2 \mu s $.}
\label{fig:BzReconst}
\end{center}
\end{figure}

A minimal compromise is to set the velocity equal to a constant value that matches the accelerated  velocity curve at some point $z_{fit} \sim z_{min}$. For instance, we have found good results by defining the position of the fit $z_{fit} \equiv \min{(\frac{3}{2}z_{min},\ 57 \mathrm{cm})}$, then defining the velocity in a piecewise fashion according to
\begin{equation}\label{eq:CTbackVzt2} 
\mathrm{v}(z,t) = \sqrt{2 \mathrm{a} (\max(z, z_{fit}) - z_{min})}
\end{equation} 
	        
Physically, a moderately large initial velocity is needed to be in agreement with the fact that there is a high rate of magnetic flux input from the external circuit, as well as an ongoing formation of new plasma by ionization of the steady flux of neutrals from the slowly closing gas valve.

The result of the velocity estimation for real CTIX data is shown in the left plot of figure \ref{fig:BzReconst}, and the resulting co-moving interpolation is shown in the right plot of figure \ref{fig:BzReconst}, the traces represent time slices of the field $B_z(z,t)$. One useful observation that is apparent with the interpolated signal is that the CT is expanding as it travels, an effect that is not obvious looking only at the raw time signals.

\section{Error analysis}\label{ErrAnalysis} We present two complementary methods of finding the order of magnitude of maximum errors for this method, in the following two subsections.

\subsection{First order velocity error}\label{FirstOrdErr}

It is reasonable to suppose that for well behaved functions there should be some input velocity field that yields the optimal reconstruction of the observable field $f(z,t)$, such that the error $|F(z,t) - f(z,t)|$  is minimized.

A simple approach is to consider an optimal trajectory $\Gamma$ that is a straight line with velocity $\mathrm{V}_{\Gamma}$. A deviation away from $\Gamma$ to some other non-optimal trajectory $\beta$, with $\mathrm{V}_{\beta} = \mathrm{V}_{\Gamma} + \delta \mathrm{V}$, will produce a quantifiable error in the interpolation.

\begin{figure}[!ht] 
\begin{center}
\includegraphics[width=.5\textwidth]{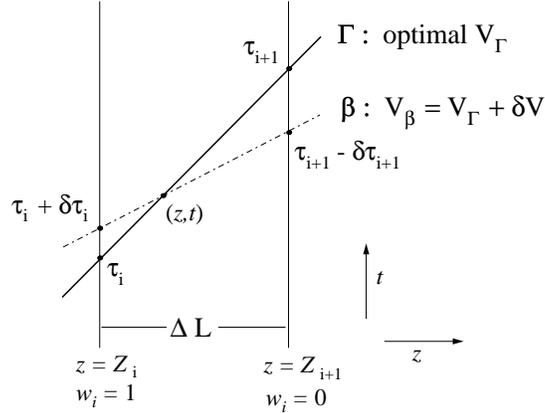}
\caption{Diagram of optimal ($\Gamma$) and non-optimal ($\beta$) trajectories for derivation of first order velocity error.} 
\label{fig:VelError}
\end{center}
\end{figure}

The difference in crossing time for the non-optimal trajectory is
\[\Delta t_{\beta}\ =\ \frac{\Delta L}{\mathrm{V}_{\Gamma}} - \frac{\Delta L}{\mathrm{V}_{\beta}}\ = \ \frac{\Delta L\ \delta \mathrm{V}}{\mathrm{V}^2_{\Gamma} + \mathrm{V}_{\Gamma} \delta \mathrm{V}}.\]

Trajectory $\beta$ will cross the probe positions at $\tau_i + \delta\tau_i$ and $\tau_{i+1} - \delta\tau_{i+1}$ where the displacements are given by
\begin{equation}
	\delta \tau_i = (1 - w_i) \Delta t_{\beta} \qquad \mbox{ and } \qquad
	\delta\tau_{i+1} = w_i \Delta t_{\beta} 
\end{equation}
 
If we evaluate the error between the optimum ($\Gamma$), and non-optimal ($\beta$) interpolations using a Taylor series for $f_i(t)$ and $f_{i+1}(t)$ about the points $\tau_i$  and $\tau_{i+1}$  respectively, formula (\ref{intfzt}) becomes 
\begin{eqnarray}
F_{\Gamma}(z,t) - F_{\beta}(z,t) &=& w_i (1 - w_i) \Delta t_{\beta}  \left(f'_i(\tau_i) - f'_{i+1}(\tau_{i+1})\right) \label{equ:BetaError}\\  
&{+}&\: \frac{w_i (1 - w_i)}{2}\Delta t_{\beta}^2 \left((1 - w_i)f''_i(\xi_i) + w_i f''_{i+1}(\xi_{i+1}) \right).{\:} \nonumber
\end{eqnarray}
The second term is a remainder that is second order in $\Delta t_{\beta}$. Equation (\ref{equ:BetaError}) holds exactly for some constants $\xi_i$ and $\xi_{i+1}$ such that $\tau_i < \xi_i < \tau_i + \delta\tau_i$, and $\tau_{i+1} - \delta\tau_{i+1} < \xi_{i+1} < \tau_{i+1}$.

We see that for fixed $\Gamma$ only the weighting function depends on position $w_i = w_i(z,t)$, and so the interpolation error is proportional to $w_i(z,t) (1 - w_i(z,t))$, which goes to zero at $z = Z_i, Z_{i+1}$ and it will have a maximum of 1/4  when $w_i = 1/2$ (at or near the midpoint $z = (Z_i + Z_{i+1})/2$).

The assumption of straight line trajectories is not fundamental, this analysis generalizes to curved trajectories yielding an identical error bound. The details of this generalization are unnecessary for this paper and contain only notational complications. Equation (\ref{equ:BetaError}) holds for any curved trajectory, the only change is that the term $\Delta t_{\beta}$ contains higher order terms in $\delta \mathrm{V}$ that do not affect the order of magnitude of the error for small $\delta \mathrm{V}$. The error is bounded by a maximum value,
\begin{equation}
\max\left(\frac{F_{\Gamma} - F_{\beta}}{F_{\Gamma}}\right) \sim 
\mathit{O}\!\!\left(\frac{\Delta L\ \delta \mathrm{V}}{\mathrm{V}^2_{\Gamma} \zeta} \right)
 = \mathit{O}\!\!\left(\frac{ L\ \delta \mathrm{V}}{(N+1)\mathrm{V}^2_{\Gamma} \zeta} \right),\label{equ:FirstOrdBound} 
\end{equation} 
where $L$ is the total length of the $z$ domain, $N$ is the number of probes used to take measurements, and $\zeta$ is the characteristic time defined in (\ref{equ:decaytime}).
The interpolation error is proportional to the local velocity error $\delta \mathrm{V}$. This result demonstrates the advantage of a high speed flow, and the use of many probes. 

The first order term of (\ref{equ:BetaError}), 
\begin{equation}
	(F_{\Gamma} - F_{\beta})_{(1)} = w_i (1 - w_i) \Delta t_{\beta}  \left(f'_i(\tau_i) - f'_{i+1}(\tau_{i+1})\right) \label{equ:F1BetaErr}
\end{equation}
can be evaluated for real data, given a velocity field, to yield upper and lower error bounds on a reconstruction. 
To demonstrate the validity of this error bound, the necessity of condition (\ref{equ:decaytime}), and the overall performance of this interpolation we have performed a more difficult, $N = 2$ interpolation using only the probe signals $B_z(57,t)$ and $B_z(142,t)$ as input data, to attempt to reconstruct the actual magnetic measurements from the $z = 91$cm probe. To optimize the reconstruction we will use the estimated velocity field that is implied by all three probe signals (57, 91, 142). 
\begin{figure}[!ht] 
\begin{center}
\includegraphics[width=\textwidth]{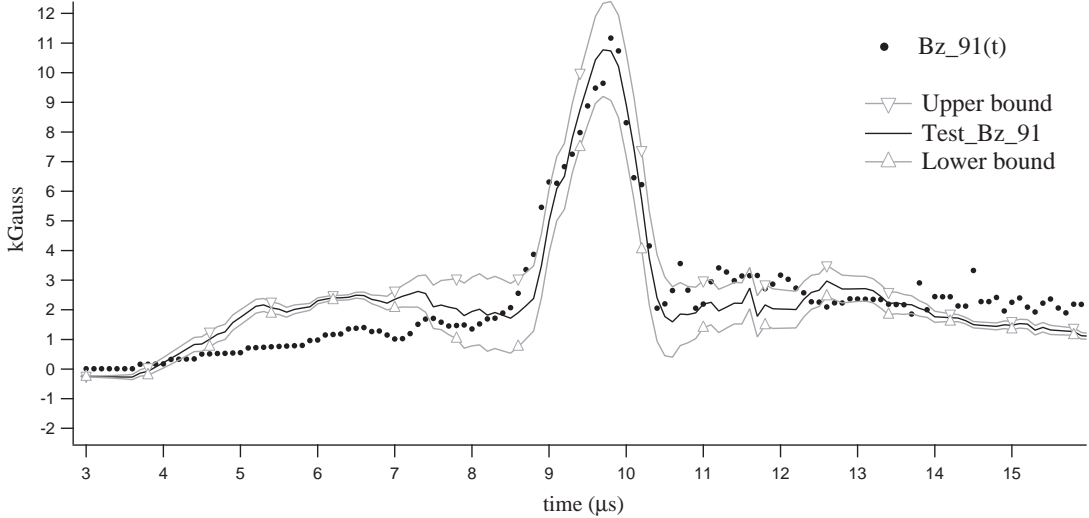}
\caption{Test comparison with real data. $N = 2$ interpolation at $z = 91$cm (--- Test\_Bz\_ 91) using only $B_z(57,t)$ and $B_z(142,t)$ as input data, compared to the measured probe signal $B_z(91,t)$ ( $ \bullet $ ) for the same shot. Equation (8.2) is used to estimate the upper and lower bounds of the reconstruction error.}  
\label{fig:Bz91comp}
\end{center}
\end{figure}

To apply (\ref{equ:F1BetaErr}) to this reconstruction we will need to assume that our velocity field is close to optimal. The estimated velocity error $\delta \mathrm{V}$ has a maximum of $5\ \mbox{cm}/\mu s$. For our probe locations the linear weighting function (\ref{equ:zweight}) yields $w_i(1 - w_i) = 0.24$. In figure \ref{fig:Bz91comp} we see the good agreement of the $N = 2$ reconstruction and the real probe signal at $z = 91$ cm.  However, within the precursor plasma at times $t = 4\ \mbox{to}\ 8\ \mu s$, the growth/decay rate condition (\ref{equ:decaytime}) is not met by this test system with only two probes. During this early time period, the relatively light precursor plasma evolves on a faster timescale than the body of the CT. The large precursor of about 4 kGauss at the $z = 57$ cm probe decays down to 1.3 kGauss by the $z = 91$ cm probe and then stays almost constant for the remainder of the acceleration, see figure \ref{fig:BzTrio59842}. This demonstrates the critical importance of condition (\ref{equ:decaytime}) in designing a probe array. For the precursor plasma, $N = 2$ is not enough; the precursor decay timescale is $2.7\ \mu s$, while the temporal span of the $N = 2$ interpolation is $4.2\ \mu s$. We see that $N = 3$ is good enough, since the span of the interpolation is $2.1\ \mu s$.

Figure \ref{fig:Bz91comp} also demonstrates that for the main structure of the CT  $(t > 8\ \mu s)$, where condition (\ref{equ:decaytime}) does hold,  the estimated error (\ref{equ:F1BetaErr}) correctly bounds the actual error in the vicinity of the compact torus. Thus, condition (\ref{equ:decaytime}) is needed both for an accurate interpolation, and also to make solid error estimates for use in subsequent data analysis of the reconstructed fields.

\subsection{Constraints on the amplitude of undetectable transient pulses} \label{subsec_Pulse}
We will now take a different approach and consider the maximum errors that could exist in the space between consecutive probes. This will include errors that are as large as mathematically allowable, with no concern initially for the physical limitations on such errors. The result of this analysis is a constraint in the form of an uncertainty relation between the duration and spatial extent of any undetected transient pulse-like modulations of the field occurring in the unmeasured region between the probes. This will provide a solid upper limit to the total error of the reconstruction. 

Since the probes themselves yield accurate and reproducible measurements of the field quantities in the immediate vicinity of the probe locations, there is little error due to global fluctuations of the real field, since these would be detected simultaneously by multiple probes. Instead, the real cause for concern are \emph{transient fluctuations}, that cause error because they are short enough in duration and spatial extent, and happen to occur deep enough into the empty space between probes so that they go undetected, and consequently the interpolation method has no ability to include their existence in the reconstructed waveform.

\begin{figure}[!ht] 
\begin{center}
\includegraphics[width=.75\textwidth]{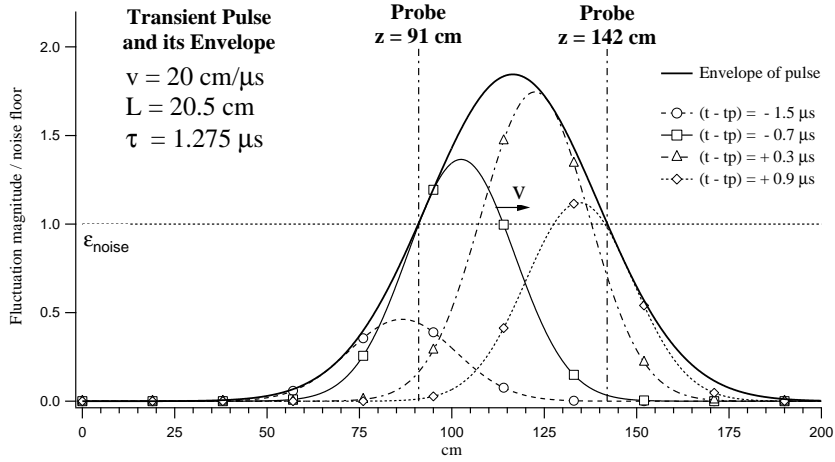}
\caption{Transient pulse at various times and its time-envelope as a function of axial position.}
\label{fig:transPulses}
\end{center}
\end{figure}

It is informative to consider transient pulses that have a Gaussian form in space and time, 
\begin{equation}
{\mathcal F}(z, t) = A\ e^{\textstyle{ -{\big(\frac{t\ -\  t_p}{\tau}\big)}^2}} e^{\textstyle{ -{\big(\frac{z\ -\  \mathrm{v} t}{\ell}\big)}^2}},
\end{equation}
where A is the amplitude of the pulse, $\tau$ its duration, $\ell$ the characteristic spatial extent, $t_p$ the time at which the pulse reaches its peak value, and  $\mathrm{v}$ is the pulse group velocity, which we will assume is approximately the average fluid velocity.  For a given probe separation $\Delta L_{probe}$ it is useful to define a dimensionless \emph{pulse extent} equal to $(\ell^2 +\mathrm{v}^2 \tau^2)/\Delta L_{probe}^2$. To measure the error introduced by this pulse in our reconstruction, we need the envelope $\mathcal{F}_{env}(z)$ of the pulse as a function of position $z$, which is the maximum value of the pulse at a fixed $z$, over all values of time. 
\begin{equation}
\mathcal{F}_{env}(z) = A \exp{\left(-\frac{\left(z -  \mathrm{v} t_p\right)^2}{\ell^2 + \mathrm{v}^2 {\tau}^2} \right)}.
\end{equation}

In order for this pulse to be undetected by neighboring probes located at $Z_i$ and $Z_{i+1}$ we would need the pulse envelope to be less than some noise floor $\varepsilon_{noise}$
\begin{equation} \label{eq:pulseineq} 
\mathcal{F}_{env}(Z_i)  < \varepsilon_{noise}  \qquad \mbox{and} \qquad \mathcal{F}_{env}(Z_{i+1})  < \varepsilon_{noise}, 
\end{equation}
where $\varepsilon_{noise}$ is the expectation value of fluctuations within an ensemble of measured signals. This will have contributions from electrical noise as well as experimental irreproducibility. It is possible that real fluctuations could occur at a larger than normal amplitude and in a transient fashion between the probes, and thus go undetected. In the worst case, the peak of the transient pulse would occur at exactly the half-way point between consecutive probes, thereby taking advantage of the largest possible amplitude allowable by ($\ref{eq:pulseineq}$). In this case, we could make the most conservative estimate for the accuracy of this reconstruction method. The worst case is when $\mathrm{v} t_p = (Z_{i+1} + Z_i)/2$, and so let  $\Delta L_{probe} = 2 (Z_{i+1} - \mathrm{v} t_p)$. Then condition for non-detection is
\begin{equation}
A \exp{\left(-{\frac{\Delta L_{probe}^2}{4 \left(\ell^2 + \mathrm{v}^2 {\tau}^2\right)} }\right)}\  < \ \varepsilon_{noise}.
\end{equation}
The maximum possible undetectable amplitude is constrained by 
\begin{equation} \label{eq:maxampformula}
A  < \varepsilon_{noise} \exp{\left(\frac{\Delta L_{probe}^2}{4 (\ell^2 + \mathrm{v}^2 {\tau}^2)}\right)}.
\end{equation}

\begin{figure}[!ht] 
\begin{center}
\includegraphics[width=.75\textwidth]{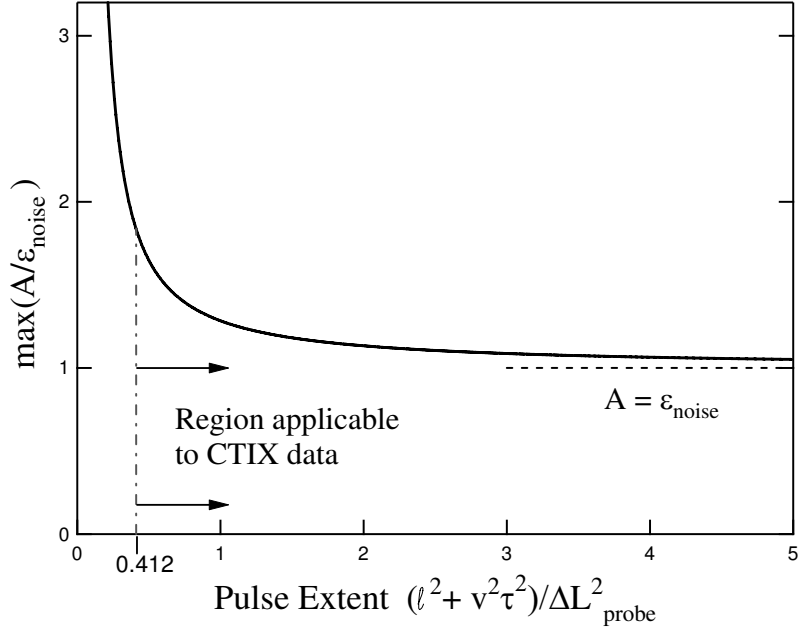}
\caption{The maximum possible amplitude of undetected transient pulses, plotted as a function of pulse extent. The maximum amplitude is determined by the inequality (8.9).}
\label{fig:transAmpMax}
\end{center}
\end{figure}

This inequality is plotted in figure \ref{fig:transAmpMax}, where the curve indicates the maximum possible amplitude for an undetectable pulse of characteristic length $\ell$ and duration $\tau$, given a probe separation of $\Delta L_{probe}$. 

We can use this result to examine the validity of co-moving interpolation being applied to the CTIX data. For an average fluid velocity of $\mathrm{v} = 20\ \mathrm{cm}/\mu s$, in the region between \mbox{$Z_i =$ 91 cm} and \mbox{$Z_{i+1} =$ 142 cm}, if there exists an undetected transient pulse that spans the space between the probes, and it exists for at least the transit time between probes \mbox{$2 \ell = 51\ \mathrm{cm} \Rightarrow \ell = 20.5\ \mathrm{cm}$} and \mbox{$2 \tau = (51/20) \mu s \Rightarrow \tau = 1.275 \mu s$} (see figure \ref{fig:transPulses}), then the pulse extent is 0.412 and the amplitude of the pulse can be no larger than 
\[A_{max} = 1.84\cdot \varepsilon_{noise}.\]

For pulses significantly shorter in duration and extent, the upper limit on amplitude becomes much larger than this example. In fact, formula (\ref{eq:maxampformula}) goes to $\infty$ in the limit of zero pulse length and duration. However, since such narrow fluctuations, if they exist, should occasionally occur directly at a probe location, their absence at the probes implies their absence between the probes as well.

If there is some physical reason for some finite upper bound on $A$, such as conservation of energy perhaps, then we can restate equation (\ref{eq:maxampformula}) in the form of an uncertainty relation

\begin{equation}
\ell^2 + \mathrm{v}^2 {\tau}^2  < \frac{\textstyle{\Delta L_{probe}}^2}{\textstyle{4 \ln{\left(A/\varepsilon_{noise}\right)}}}. \label{equ:Uncertainty}
\end{equation}

For some fixed $A > \varepsilon_{noise}$, there is a trade-off between the spatial extent and the duration of undetectable pulses, and that this relationship depends on the fluid velocity and the distance between consecutive probes. In agreement with the result from section \ref{FirstOrdErr}, inequality (\ref{equ:Uncertainty}) shows that the best reconstruction occurs when consecutive probes are closely spaced, and there is a high flow velocity. 

\section{Conclusion}
Co-moving interpolation is a method that is applicable to many experimental and industrial problems where the properties of a fluid flow are being measured at several discrete locations, and the global dynamics of the system need to be determined. The method is straightforwardly applied, and makes efficient use of modern computer power. High accuracy can be achieved with a minimal number of probe points if conditions such as (\ref{equ:decaytime}), (\ref{equ:FirstOrdBound}) and (\ref{equ:Uncertainty}) are employed in the design of diagnostic systems.

We have presented the details of a efficient algorithm for this method, which we have implemented on a desktop computer.  We have run our method on a sequence of test problems in which an analytic model of shock propagation is compared to reconstructions by co-moving interpolation. These tests demonstrated the need for an optimal velocity field, its distinction from the fluid velocity field when diffusion is present, as well as the high degree of accuracy that is possible when the optimal velocity field is used.

We have applied our method on the CTIX system, and presented the resulting co-moving interpolation of $B_z(z,t)$. In future articles we will apply this method extensively to the CTIX plasma system to investigate MHD effects during the acceleration process.\cite{DavisThesis} In addition, the results of this method are allowing a new kind of comparison between experimental data and ongoing MHD simulations of the CTIX system. Ordinarily, multi-dimensional computational studies of physical systems produce results that can be directly visualized, or to make comparisons to experiment the simulations can be sub-sampled to produce virtual diagnostic signals, which are directly compared to raw diagnostic signals from the corresponding experiment. An alternative method of comparison is to start with a set of raw experimental data from multiple diagnostics, and then apply a method of analysis to synthesize the approximate global behavior of the system, and then make the comparison to simulation. Both routes have advantages, however, at present there are few tools to solve the inverse problem on the experimental data to make use of the second route of comparison. The combination of animated visualization of experimental probe data, and quantitative analysis on the reconstructed fields, is providing new insight into the nature of the magnetized plasma in CTIX.

Lastly, we considered two methods for making error bounds on this interpolation. In the first analysis we found good agreement of the first order velocity error (\ref{equ:F1BetaErr}) with the actual error of the $N = 2$ reconstruction of $B_z(91,t)$ signal from CTIX. The second analysis showed that for the conditions present on CTIX, undetectable transient pulses of significant size and duration would be limited to a level that was within a factor of 2 of the noise floor of the probes. Both methods derived similar relationships between flow velocity, probe separation and the resulting error of the interpolation.

\end{document}